\documentclass[letterpaper,twocolumn,10pt]{article}
\usepackage{usenix2025_SOUPS}
\usepackage[table]{xcolor}
\usepackage{csquotes}
\usepackage{enumitem}
\usepackage{graphicx}
\usepackage{tcolorbox}
\usepackage{booktabs}
\usepackage{multirow}
\usepackage{adjustbox}
\usepackage{wrapfig}
\usepackage{picinpar}
\usepackage[utf8]{inputenc}
\usepackage{caption}
\usepackage{cuted}
\usepackage{authblk}

\usepackage[hyphens]{xurl}
\definecolor{rowcolor}{HTML}{FFC20A}

\newcommand{\intq}[2]{$#1#2$}

\renewcommand{\paragraph}[1]{\vspace{0.1in}\noindent\textbf{#1}}

\makeatletter
\def\@listi{\leftmargin\leftmargini
    \parsep 1\p@ \@plus0\p@ \@minus\p@
    \topsep 2\p@   \@plus0\p@ \@minus\p@
    \itemsep1.25\p@ \@plus\p@ \@minus\p@}
\let\@listI\@listi\@listi
\makeatother

\begin{document}

\title{``The System Will Choose Security Over Humanity Every Time'': Understanding Security and Privacy for U.S. Incarcerated Users 
}

\def\plainauthor{Yael Eiger}
\def\plainauthor{Nino Migineishvili}
\def\plainauthor{Emi Yoshikawa}
\def\plainauthor{Liza Nadtochiy}
\def\plainauthor{Kentrell Owens}
\def\plainauthor{Franziska Roesner}

\author[1]{Yael Eiger}
\author[1]{Nino Migineishvili}
\author[1]{Emi Yoshikawa}
\author[1]{Liza Nadtochiy}
\author[2]{Kentrell Owens}
\author[1]{Franziska Roesner}
\affil[1]{University of Washington}
\affil[2]{Max Planck Institute for Security and Privacy}
\maketitle

\begin{abstract}
  Digital devices like tablets, media players, and kiosks are increasingly deployed in U.S. prisons. These technologies can enable incarcerated people to access education, communicate with loved ones, and develop vital reentry skills. 
 
  However, they can also introduce new privacy and security risks for incarcerated people who have little agency over their usage and contracts, and are currently carved out of many consumer protection safeguards.
  To investigate these issues, we conducted focus groups and interviews with system-impacted people (n=17), i.e., those formerly incarcerated, and their relatives, to investigate experiences with device-related security and privacy vulnerabilities and the power dynamics that affect their use. In our findings, participants describe pervasive surveillance, censorship, and usability problems with the technology available to them, including shifting and seemingly arbitrary usage policies. 
  
  These policies strain relationships both inside and outside prisons and contribute to negative downstream effects for incarcerated users. 
  We recommend ways to better balance prison security concerns with privacy-related needs of system-impacted individuals by promoting accountability for technology-related decisions, providing public oversight of digital purchasing and use policies, and designing digital tools with them---the actual end-users---in mind.
\end{abstract}

\section{Introduction}

Tablets, media players, kiosks, and other digital devices are proliferating throughout United States prisons and jails~\cite{Surrett_Corona_Weill-Greenberg_Samuels_Bonds, Weill-Greenberg_Ballesteros_O’Connor, West_2023, Bertram_Wagner}, enabling incarcerated people to access education and legal services, communicate with loved ones, and learn vital skills for re-entering our increasingly digital society~\cite{studyingup, ertl2019social, reisdorf2022digital}. The use of  technology in this context is extremely common and far reaching, directly affecting an estimated 1.9 million currently incarcerated people in the U.S. and another 113 million people (45\% of the U.S. population) who have an immediate family member who has been to jail or prison~\cite{prisonpolicyinitiative}. Collectively, this cohort is defined by the UC Berkeley Underground Scholars Initiative as \textit{``system-impacted''} people~\cite{czifra2022analyzing}.

Prior research has documented security and privacy issues that arise due to other technologies in the carceral system, such as electronic monitoring for people on parole or probation~\cite{owens22usenixsecurity}. Other work has studied experiences with technology post-release, such as the experiences of formerly incarcerated people who are navigating their release from prison and re-entering society~\cite{reisdorf2022digital, ogbonnaya2018returning, ogbonnaya2019towards, seo2022returning}. One prior work at CHI studied the experiences with surveillance of loved ones communicating with their incarcerated relatives\cite{kentrellCHI}. In this work, we take a broader perspective, focusing on the experiences of people who directly experienced technology while incarcerated, including technologies they had access to beyond communication.

Closing the public knowledge gap about experiences \textit{while} incarcerated is crucial because incarcerated people have limited agency regarding when, how, and whether they can use digital devices~\cite{arguelles2021bars, Bertram_2019}. Their access is determined and restricted by specific prison policies and contracts with digital system purveyors. Further, two companies, ViaPath and Securus, dominate the market supplying prisons and jails with digital devices~\cite{Floberg_McDermott_2024}. 
Such limited options have been shown to burden and harm a disadvantaged and vulnerable population who already lack legal rights to privacy and the safeguards provided by consumer protection laws~\cite{arguelles2021bars, Bertram_2019, Littman_2022}. 

Moreover, the policies that govern the use of the digital devices allowed in the carceral system are often constrained by concerns about the  ``security'' risks they pose for the prisons~\cite{knkxWashingtonExperiments, Wright_1996, oclcComputingBehind, nytimesLawInmates, West_2024}. 
Examples from the 1980s and 1990s, through more recent ones from 2024, show that prison officials often cite unsubstantiated security threats to justify revoking computer access; this is true even when programs demonstrate clear rehabilitative benefits and near-zero recidivism~\cite{Wright_1996, oclcComputingBehind, nytimesLawInmates}.

In this work, we investigate the security and privacy experiences of system-impacted people using digital devices while incarcerated, from their own perspectives. We investigate three questions:
\begin{itemize}
    \item \textbf{RQ1:} What security and privacy threats, risks, and concerns do system-impacted people experience when \textbf{utilizing digital devices while incarcerated?}  
    \item \textbf{RQ2:} How do system-impacted people perceive security and privacy-related \textbf{prison policies and their implementation} by prison officials?
    \item \textbf{RQ3:} What 
    \textbf{downstream effects} do system-impacted people experience as a result of these issues?
\end{itemize}

To that end, we conducted focus groups and interviews with $17$ system-impacted participants, including 12 formerly incarcerated people and 5 loved ones.

Our participants reported that they experienced pervasive forms of surveillance and censorship related to available technologies during incarceration. They described shifting and arbitrary technology censorship policies that strained their relationships with outside communities and threaten their future rehabilitation. They related experiences of living in a ``panopticon''\footnote{The term ``panopticon'' was popularized by Foucault \cite{foucault_discipline_1979}, but the ``panopticon'' prison was originally designed by Bentham. The panopticon guarantees that incarcerated people are unaware of when or whether they are being watched at a given moment, which in turn creates self-regulation. } 
of surveillance that exacerbates feelings of dehumanization, self-doubt, and mistrust towards technology and the prison infrastructure. This lack of privacy further caused participants to feel at once powerless to affect and grateful for the scant access to technology they do have, described by one  participant as ``technology Stockholm syndrome.''\footnote{The term our participant used, ``Stockholm Syndrome,'' refers to the phenomenon observed in Stockholm, Sweden of a hostage developing a bond with or even gratitude towards their captor, which has also been described as ``a conscious coping strategy that can be understood as a form of adaptive behavior, providing hope for the victim in an otherwise hopeless situation''~\cite{JamesonCeliaStockholmSyndrome}.}  

Our participants also described how prison policies regarding technology use stem, in their view, from incomplete or misguided mental models that prison officials hold about the incarcerated population and their ability to misuse technologies, as well as the technologies themselves. The concerns about prison security raised by these mental models lead in part to the technology restrictions and surveillance (and the downstream harms) that our participants describe. 
Beyond giving voice to the experiences of our participants, our findings raise important considerations concerning ways in which normative security and privacy beliefs become replicated and amplified through the medium of digital technologies. Presumptive security and privacy concerns expressed by prison administration can give rise to a very real, practical lack of security and privacy for incarcerated people and their loved ones. This finding provokes and aligns with broader questions asked within recent security and privacy research, namely, the consideration of marginalized and/or vulnerable populations in threat modeling~\cite{heilmeier_article}. In other words: security and privacy \textit{for whom}?

\section{Background, Related Work, and Motivation}\label{sec:background}

\subsection{Digital Devices in U.S. Prisons}
ViaPath and Securus build and distribute digital devices such as tablets and media players to prisons. These devices come pre-loaded with their messaging applications ``GettingOut''/GTL for Viapath and ``JPay,'' which became Securus after a 2015 buyout ~\cite{jpayAquisition2015}. To send messages, loved ones of incarcerated people (outside of prison) separately access the GettingOut or JPay/Securus applications by downloading them on their personal devices. 
The Prison Policy Initiative estimates that these two companies control more than 80\% of the market share for carceral communication technology~\cite{Floberg_McDermott_2024, prisonpolicySMHRapid}, operating in thousands of facilities~\cite{securustechAboutSecurus,viapathAboutViaPath}. In our geographic area, the prisons predominantly contract with Securus; thus, our participants had only used JPay and Securus, and not Viapath products. 

\begin{figure}
    \centering
    \includegraphics[width=\linewidth]{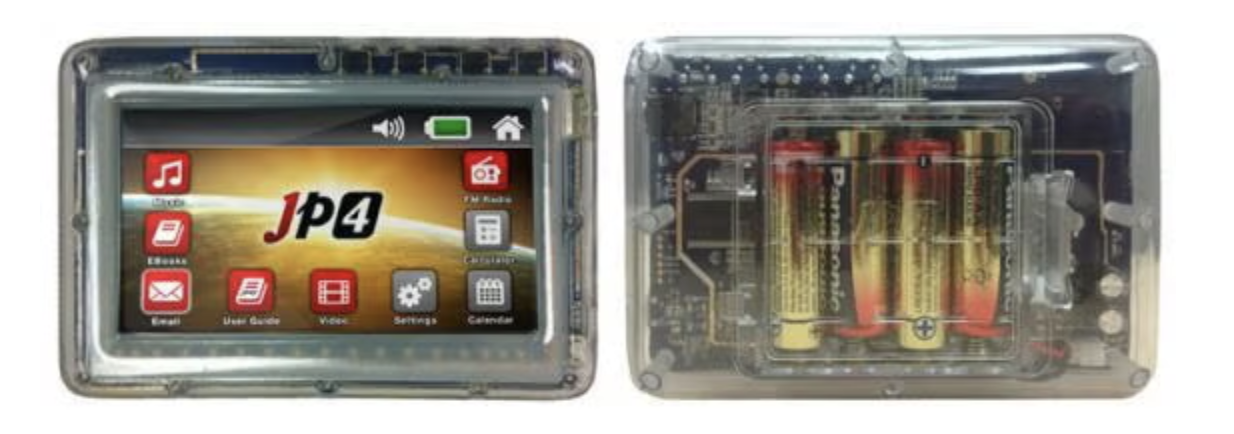}
    \caption{An example tablet, the JPay4, that might be available to an incarcerated person. The tablet's case is transparent, and it features apps for messaging, listening to music, or accessing legal services.}
    \label{fig:JPAY_tablet}
\end{figure}

These technologies and apps are changing rapidly. Over the past decade, for example, telephones have been almost wholly replaced by kiosks (wall-mounted computer screens protected by a steel box)~\cite{Koebler_2024}. Kiosks, in turn, are now being supplemented by personal media devices and tablets ~\cite{Koebler_2024}, as shown in Figure \ref{fig:JPAY_tablet}. Such devices are used for communication purposes (e.g., messaging), though our participants also described other purposes, such as using the (same) tablets as music/media players and to access legal services.

\subsection{Security and Privacy Concerns for System-Impacted People}

\paragraph{During Incarceration.}
The privacy and security concerns with technologies available within prisons are not hypothetical. 
ViaPath and Securus have previously come under scrutiny for poor privacy practices and the increased surveillance of users. 

For example, in 2018, Securus was sued for collecting and storing location data, audio recordings, and transcripts of phone calls from people in prison, including phone calls protected by attorney-client privilege~\cite{Valentino-devries_2018, Lee_Smith_2015}. Securus settled the lawsuit and their contract was again renewed with the correctional facility~\cite{Fry_2018, Smith_2016}. 

Chilling effects and privacy concerns due to surveillance, as well as lack of technology access, can have downstream impacts. For example,  connection to community support and legal counsel is being increasingly mediated (and restricted) through digital technologies~\cite{jewkes2016brave}. However, access to one's community has been shown to improve mental health and wellness, reduce recidivism, and disciplinary action during incarceration~\cite{Caplan_2024}. 
Prior work has also explored how the lack of educational access inside prison leads to a stark digital divide when they reenter society \cite{ogbonnaya-ogburu_nonprofits,ogbonnaya2018returning, ogbonnaya2019towards, reisdorf2018digital, reisdorf2022digital, grierson2022design, anuyah2023characterizing, gautam2024enhancing, nisser2024prisons, martinez2024engaging}. 

Thus, security, privacy, and related concerns for the system-impacted population exist. 
But to our knowledge, their own self-reported experiences, and the downstream effects of those concerns,
have been understudied: for example, Verbaan et al.~\cite{verbaan2018potentials} notes that the prison context has been largely 
overlooked in computer science, especially with respect to how system-impacted people experience and use technology.   

Previous academic work by Owens et al.~\cite{kentrellCHI} considered \textit{the families and loved ones} of currently incarcerated people and their mental models of surveillance when using carceral communication tools. Our findings support and strengthen those of this prior work, as well as add new perspectives.
Specifically, we gather the perspectives of \textit{formerly incarcerated people themselves} (in addition to their families or loved ones),
and we consider technology use cases beyond communication.
Our findings also extend beyond surveillance to other privacy, usability, and consumer protection concerns, as well as downstream impacts.

\paragraph{After Release.}
After release from prison, systems-impacted people continue to experience technology restrictions and associated security and privacy impacts. 
For example, Owens et al.~\cite{owens22usenixsecurity, owensFAccT2025} investigated the technical, legal, and human-centered impacts of electronic monitoring applications used for parole, probation, and immigration restriction. 

Prior work on digital technologies post-release also shows that privacy concerns can lead formerly incarcerated people to avoid or limit their engagement with digital technologies, and many people cite surveillance concerns in particular~\cite{ogbonnaya2018returning, seo2022returning, kentrellCHI}. For example, Seo et al.~\cite{seo2022returning} examine women transitioning from incarceration and note that post-incarceration discomfort with the internet often stems from its association with the surveillance they experienced in prison. As such, online privacy concerns often make people, in particular underserved populations, hesitant to acquire new skills and engage with technology~\cite{seo2019evidence}. Gurasami et al.\ ~\cite{gurusamiWeb} coined the term ``carceral web'' to refer to a lack of agency formerly incarcerated people feel over their privacy, due to the internet's ability to publicize, archive, and document their crimes.

\subsection{Notions of ``Security'' in Carceral Systems}

Prisons have a long history of incorporating and then revoking technology access for the incarcerated. For example, in 1986, incarcerated people at the Washington State Prison in Monroe were allowed to purchase computers and other devices to keep in their cells~\cite{Wright_1996}. According to Paul Wright, founder and director of the Human Rights Defense Center, recidivism rates for these people was ``nearly zero,'' and many found employment from the computer skills they acquired even after prolonged prison sentences~\cite{Wright_1996, nytimesLawInmates, oclcComputingBehind}. A few years later, a new director terminated the program, citing it as a ``security issue,'' even though he acknowledged under oath that no threat had arisen due to the program~\cite{Wright_1996, nytimesLawInmates}. 

The policies that govern the use, and often revocation, of digital devices continue to be justified with notions of ``security'' risks~\cite{knkxWashingtonExperiments, Wright_1996, oclcComputingBehind, nytimesLawInmates, West_2024} for the prisons. As recently as 2024, officials restricted access to the ironically named ``Securebook 5'' computers for incarcerated students based on amorphous security concerns~\cite{West_2024}.

In the prison context, the considerations of ``security'' are inextricably linked with the practices of privacy. For instance, a new AI system called Verus was deployed across correctional facilities in eight states to monitor calls between incarcerated people and their loved ones; the system monitored 2.5 million calls during a 2021 year-long pilot in Suffolk County, New York alone~\cite{Asher-Schapiro_Sherfinski_2021}. The stakeholders, this time governing the introduction of a new technology that accentuated surveillance while diminishing user privacy, were once again correctional facilities aiming to detect ``security threats''~\cite{Asher-Schapiro_Sherfinski_2021}.

These digital technologies are designed with prisons as the main stakeholders. 
Prisons hold the authority to determine how to implement and deploy technologies within their facilities based on their unique security and privacy conceptions. Their motivations and viewpoints are well-documented in prior work~\cite{mufarreh2022, lerman_2012, Vachiradath01102013}.
In contrast, our paper focuses on a different stakeholder: system-impacted people as the direct end-users of these technologies. For instance, when looking at the Verus system 
from the perspective of the actual end-user, we see privacy breaches: 

Verus flagged conversations of system-impacted people referencing the Spanish word for ``lawyer'' and conversations concerning COVID-19 cover-ups in detention facilities~\cite{Asher-Schapiro_Sherfinski_2021}. Viewing Verus from the perspective of the incarcerated reveals a different notion of ``security and privacy,'' one where safety, health and right to confidential communication with legal counsel are jeopardized.

\section{Methodology}\label{sec:methods}

We aimed to uncover (i) the security and privacy concerns of system-impacted people while incarcerated, (ii) their perceptions of security and privacy related prison policies, as well as (iii) the downstream effects of said concerns.

\subsection{Positionality}
As researchers, our experiences and worldviews alter the perspectives we bring to our work~\cite{wei2024sokorsolkquantitative, fourtensions}. Thus, we describe our identities and positionality here. The research team included people with varied ethnicities and races, socioeconomic, geographic, and educational backgrounds, as well as a diverse set of experiences with policing and the carceral system. Multiple authors have spent years teaching classes in prisons and working closely with re-entry organizations. Significantly impacting our perspective, however, is the fact that no researcher on our team has personally been incarcerated.

\subsection{Recruitment} \label{ss:recruitment}

Participants were recruited through advocacy organizations that we have community partnerships with, such as [anonymized for review]. These organizations are led by, include, and maintain established, trusted relationships with the communities we sought to engage. In addition to supporting recruitment, these partnerships provided a foundation for more sustained support for participants beyond the study itself and offered a continuity of care and resources rather than a one-time interaction. The authors also continue to nurture and give back to these community partnerships, subsequently co-organizing and co-designing workshops and events together. 

After identifying and enrolling the first few participants through these organizations, we employed snowball sampling~\cite{snowballSampling,biernacki1981snowball} to expand our participant pool. This method leveraged participants' existing social networks.

Our fliers asked if participants had any experience with carceral technology such as JPay or Securus, but it did not reveal other research questions we planned to study.

Participants were subsequently asked to fill out a screening survey, which we used to verify their eligibility. Participants had to be over 18, have firsthand experience using JPay/Securus or GettingOut, and be formerly incarcerated or loved one of someone who is currently incarcerated. We chose not to directly interview currently incarcerated people to minimize potential risks to that population (see Section~\ref{sec:limitations}).

For confidentiality reasons, we chose not to collect or report demographics, including race, at the individual participant level. However, race has a strong impact on over-policing and incarceration rates in the U.S.~\cite{bales2012racial, sentencingprojectFiveDisparities, ppicRacialDisparities}. Black people have an incarceration rate 4-6 times higher than white people, Latinx people have an incarceration rate 3-4 times higher, and Native American people have an incarceration rate that is twice as high as white people ~\cite{nativeAmericansPrisonPolicy,warren2012imprisonment, Ghandnoosh, sentencingprojectFiveDisparities, ppicRacialDisparities}. To ensure we captured these experiences but did not risk participant privacy, we intentionally recruited from organizations [anonymized for review] that represent and advocate for these historically over-incarcerated identity groups, including an advocacy group for incarcerated Black people, an advocacy group for incarcerated Native American \& Indigenous people, and a group that helps recently released people struggling with housing insecurity.  

\subsection{Participants} \label{ss:participants}

In total, we conducted 4 focus groups and 5 one-on-one interviews, with a total of 17 participants. 
We offered the option of focus groups to all participants, in part because prior research indicates that vulnerable groups  may feel more comfortable sharing their experiences in group settings~\cite{race1994rehabilitation, pollack2003focus} and 
when participants are already acquainted. 
Thus, based on participant preferences, we conducted some focus groups among participants who already
knew each other (i.e., with their family members or social connections that were also eligible for our study)
as well as some individual interviews. Spanning these focus groups and interviews, twelve participants were formerly incarcerated, and five were family members of currently incarcerated people or formerly incarcerated people (see Table \ref{tab:participants}). Specifically, the family member in Focus Group 2 (F1) is both family to their currently incarcerated spouse, and family to the formerly incarcerated people in their focus group (P4, P5). The Family member in Interview 1 (F2) has a currently incarcerated spouse. The family members of Focus Group 4 (F3, F4, F5) are family members of the formerly incarcerated participant in their focus group (P10). 

We iteratively analyzed data as we collected it, described below.
We reached thematic saturation, i.e., no new themes emerged, after 11 participants, although we continued with our scheduled 
focus groups and interviews until we reached 17 total participants. 

\begin{table}[tb]
\centering
\footnotesize
\begin{tabular}{ m{0.25cm} ll m{2.4cm} }
\toprule 
~ & Session & Status & Access To \\
\midrule 
P1 & Focus Group 1 & Formerly Incarcerated & JPay, Kiosk \\
P2 & Focus Group 1 & Formerly Incarcerated & JPay, Kiosk \\
P3 & Focus Group 1 & Formerly Incarcerated & JPay, Kiosk \\

P4 & Focus Group 2 & Formerly Incarcerated & JPay, Kiosk, Securus \\
P5 & Focus Group 2 & Formerly Incarcerated & JPay, Kiosk, Securus \\

\rowcolor{rowcolor} F1 & Focus Group 2 & Family member& JPay, Securus \\
P6 & Focus Group 3 & Formerly Incarcerated &  JPay, Kiosk\\ 
P7 & Focus Group 3 & Formerly Incarcerated &  JPay, Kiosk\\

\rowcolor{rowcolor} F2 & Interview 1 &  Family member& JPay, Securus \\

P8 & Interview 2 & Formerly Incarcerated & JPay, Kiosk, Securus \\
P9 & Interview 3 & Formerly Incarcerated & JPay, Kiosk, Securus\\ 

\rowcolor{rowcolor} F3 & Focus Group 4 & Family member*& JPay, Securus \\
\rowcolor{rowcolor} F4 & Focus Group 4 & Family member*& JPay, Securus \\
\rowcolor{rowcolor} F5 & Focus Group 4 & Family member*& JPay, Securus \\
P10 & Focus Group 4 & Formerly Incarcerated &  JPay, Kiosk, Securus\\ 

P11 & Interview 4 & Formerly Incarcerated & JPay, Kiosk\\ 
P12 & Interview 5 & Formerly Incarcerated & JPay, Kiosk, Securus\\

\bottomrule 
\end{tabular}
\caption{
\textbf{Focus Group and Interview Participants.} In total, we recruited 17 participants, 12 of whom are formerly incarcerated (denoted by the letter ``P'' followed by the participant number) and 5 of whom were family members of currently or formerly incarcerated individuals (denoted by the letter ``F'' followed by the participant number, and marked visually with a \colorbox{rowcolor}{colored} background). Family members of formerly (rather than currently) incarcerated individuals are denoted with an asterisk. Participants' access to technologies depended on when and where they were incarcerated. 
} 
\label{tab:participants}
\end{table}

\subsection{Interview Logistics and Protocol}

Interviews lasted approximately one hour and were conducted either in-person or remotely, depending on the participant’s preference and comfort level. While we reserved private rooms on campus for in-person interviews, we recognized that some participants might feel uneasy or uncomfortable coming to a university setting. Additionally, we acknowledged that many participants live in rural or suburban areas where the commute to the university can be prohibitively long or expensive. To reduce barriers to participation and increase accessibility, we offered to meet participants at community-based locations such as public libraries or community centers closer to where they live.

Participants were compensated \$35 USD per hour in a prepaid VISA gift card in addition to commuting and parking costs.

Participants gave informed consent twice: (1) before scheduling interviews/focus groups, and (2) verbally again at the start of the session. The sessions were audio recorded with participants' consent.

Our interview protocol is provided in Appendix~\ref{appendix:protocol}. After establishing that everyone was comfortable, and asking some icebreaker, intro, and rapport building questions, we asked participants about their in-prison or outside experience with technology. Since different prisons contracted with different technology vendors, we asked them to clarify which technologies (e.g., JPay, Securus, GTL) they used and when. We also asked about other technology used in the prison. We then prompted participants to discuss their security and privacy concerns and what factors influenced their experiences. We asked how these practices impacted relationships and dynamics inside and outside the prison. We closed by asking participants to share their aspirations for change. 

Because we believed the interviews might contain sensitive information, we manually transcribed the recordings rather than using a third-party transcription service, stored all transcripts in directories only viewable by the research team, and used encrypted messaging tools to schedule the interviews. 

\subsection{Ethical Considerations}\label{sec:ethics}

Our study was approved by our Institutional Review Board (IRB). Given the sensitive nature of incarceration,  we took proactive measures to ensure that the research environment remained kind, safe, and participant-centered. 

We collected identifiable information including participants' full name, email address, and phone number solely to administer participant compensation, as mandated by our institution. This information was stored in a separate spreadsheet not linked to research data, where participants were identified only by unique identifiers (e.g., ``P1''). 

All data was stored in a password-protected Google Drive only accessible to the research team. Each participant was also assigned a unique participant identifier (e.g., ``P1''), which was used throughout the research documentation and analysis. Audio recordings were transcribed fully manually and no third-party service was used. All quoted material was de-identified.

We did not ask participants about their criminal charges, sentencing, or details of their incarceration. 
Anticipating that some topics might provoke discomfort, we designed interviews with built-in breaks and flexibility. 
While no participants chose to pause or end the interview early, we remained vigilant and prepared to accommodate such needs.
Participants were reminded throughout that they could skip questions or retroactively request the removal of content. 
The final manuscript was sent to participants via email two weeks before submission, so that they could highlight any areas they felt misrepresented their experiences or felt sensitive. The feedback we received from participants was positive and no one requested a removal of their quotes.

\subsection{Data Analysis}
\label{ss:dataanalysis}
Data analysis was conducted in multiple stages. Initially, we used a collaborative, qualitative open coding technique~\cite{thomas2006general}, where three researchers separately examined two different interviews and/or focus groups and drew up initial codes. These codes were then combined and further refined collaboratively. 
This led to the creation of a cotook that encapsulated codes relating to (1)~technologies used by system-impacted people and (2)~the benefits and harms of these technologies. We repeated this process with two additional interviews and/or focus groups by two researchers. This resulted in the creation of a new code addressing (3) prison policies, people or parties that govern the use of technologies, and security \& privacy concerns. Finally, two researchers applied the established codebook to all transcripts. Our final codebook is in Appendix~\ref{appendix:codebook}.
We then used inductive thematic analysis~\cite{braun2006using} to uncover overarching themes. Three authors independently derived themes from the transcripts. The themes were later converged through open discussion among all researchers.

\subsection{Limitations}\label{sec:limitations}

Our study focused on persons incarcerated in state and federal prisons in The Puget Sound region to surface findings in a context that we understood well. As such, our results should not (and, like much qualitative research, are not intended to) be interpreted as generalizing beyond this. 
We focused on technology that people use while incarcerated, rather than things like prison surveillance cameras, which incarcerated people do not use but are subjected to. 

We also acknowledge the limitation in recruiting via snowball sampling and that people who know each other may hold similar views that are not generalizable at scale. 

To minimize the amount of personal information collected, we did not gather individual demographic data (e.g., race, age, ethnicity, sentence duration, sentence dates, etc.). As a result, we are unable to analyze our findings in relation to these demographic factors. 

We also acknowledge the limitation of focusing only on formerly incarcerated people and their loved ones while excluding currently incarcerated people. Though it is possible to do research ethically with currently incarcerated people, we recognize that working with a more vulnerable group carries additional risks that we did not need to impose to answer our research questions. Following best practice, we thus interviewed proxies for this vulnerable population~\cite{belliniSOK}.

Moreover, we did not include many other key stakeholders in the carceral system. For instance, we did not interview technology developers, Department of Corrections officials and staff, technology company representatives, police officers, or lawyers. We felt there was a gap in understanding the perspectives of system-impacted people specifically, while the opinions and stances of prison officials regarding technology have been well-studied and documented~\cite{rand, statedept, National_Institute_Justice_1996}. Our choice to study the opinions of a marginalized population is in line with prior work that focuses on electronic monitoring subjects~\cite{owens22usenixsecurity}, loan app users~\cite{10.1145/3715275.3732057}, undocumented immigrants~\cite{guberek}, or survivors of stalkerware~\cite{10.1145/3173574.3174241} without including the viewpoints of technology developers and deployers.

\section{Findings} \label{sec:findingsTechnology}

\begin{figure}[tb]
\centering
\includegraphics[width=\columnwidth]{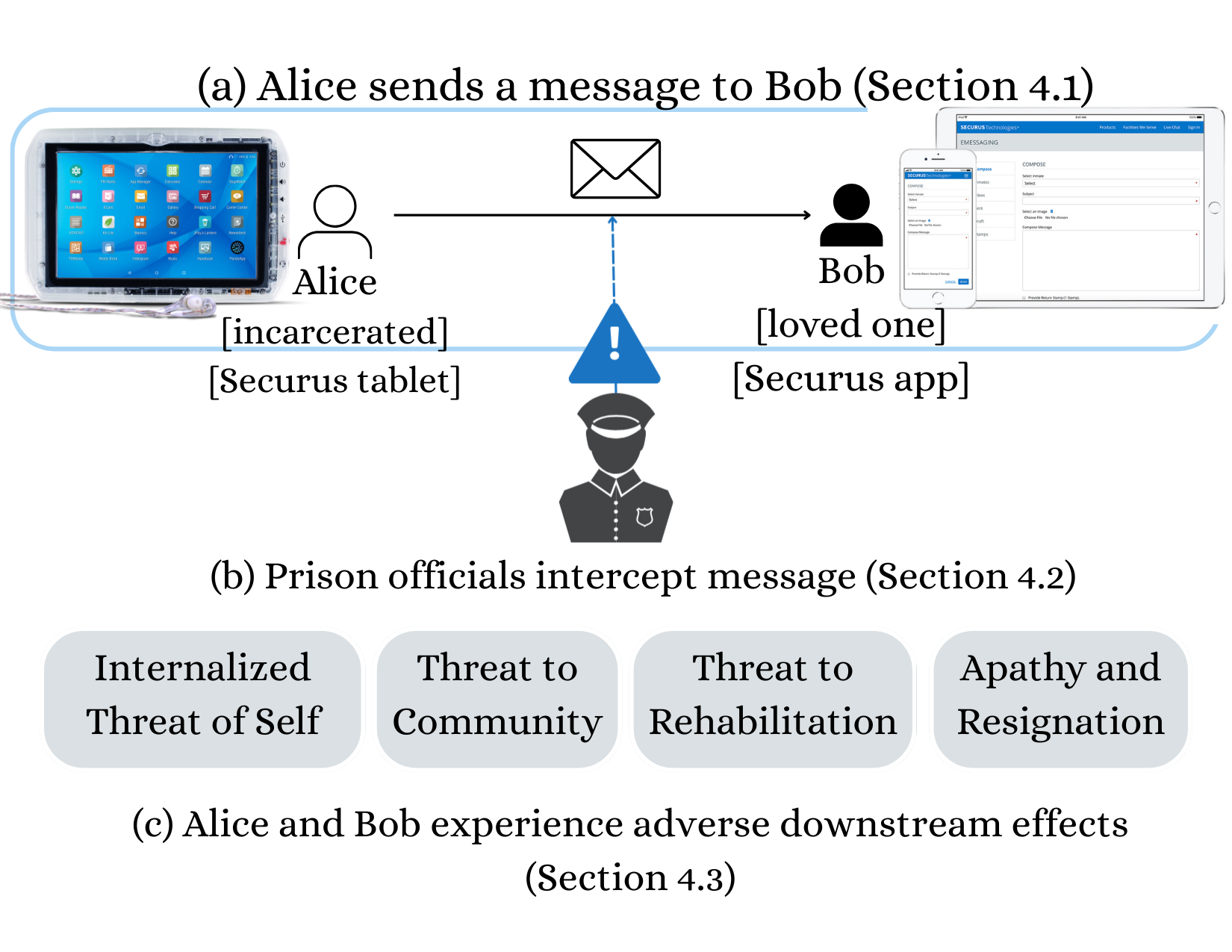}
\caption{A summary of our findings, using Alice and Bob. In (a) Alice uses a Securus tablet inside of prison to send a message to Bob, outside of prison, who uses the Securus app on his mobile phone or tablet. In (b), a prison official intercepts the message. In (c), we allude to the adverse downstream effects of this commmunication process.}
\label{fig:alicebob}
\end{figure}

We structure our findings using a version of the 
Alice and Bob  communication model often used in computer security and privacy, shown in Figure \ref{fig:alicebob}. Here, Alice is an incarcerated person attempting to communicate via prison communications technologies (e.g., Securus), and Bob is a loved one outside of prison who uses his regular mobile device. 

We also highlight the role of a third actor---the interceptor---embodied by prison officials and staff, whose responsibility it is to enforce prison policies. Throughout our findings, we detail how the role of the interceptor is to surveil, censor, and control the message flow per institutional logic, as shown in Figure \ref{fig:alicebob}b.

This then drives downstream effects for Alice and Bob.

\subsection{Direct Security and Privacy Experiences of System-Impacted People}\label{subsec:rq1}

In the initial stage of the Alice and Bob communication model, Alice attempts to send a message to Bob using prison-provided communication tools. This section describes the security and privacy concerns that emerge for Alice and Bob in this direct communication line. Participants described this step as rife with involuntary and stringent censorship (Sections \ref{ssc:censorall} to \ref{ssc:flux}) that delay communication (Section \ref{ssc:seamless}).

\subsubsection{Censorship Is Overwhelming} \label{ssc:censorall} All participants commented that censorship via digital devices is prevalent, which F2 summarized as: ``Everything is monitored. Everything is censored.'' The censorship extends both to the person that a system-impacted individual is allowed to communicate with and to the words and messages allowed to be exchanged between them. P1 summarized:
\begin{displayquote}
    ``There’s so much censorship around all of that. Right, because you’re not allowed to talk to other prisoners. You’re censored in that\ldots Even with the people that you do talk with, you guys could talk\ldots but only about certain things.'' \textit{---P1}
\end{displayquote}

\noindent Participants perceived censorship as extending beyond communication, attributing limited availability of educational materials on the devices to deliberate censorship decisions:
\begin{displayquote}
``They censor the books that we’re able to get in\ldots They censor the music. They censor the motivational speeches.''
\textit{---P1}
\end{displayquote}

\noindent The apparatus that enables this vast censorship is surveillance. P6 explained that ``[the Department of Corrections, DOC] tracks everything. Everything gets recorded.'' P9 further perceived that underpinning censorship was a system of ``surveillance of the video call and text system; every message was `reviewed' by somebody.''
This constant, unremitting prevalence of surveillance and censorship had downstream effects, as we describe in Section~\ref{subsec:adverse}.

\subsubsection{Consent Is Involuntary}\label{ssc:consentno} 
Participants reported that surveillance and censorship  breached their privacy. As F2 observed:

\begin{displayquote}
``When you think about it, it really is an invasion of privacy. Because before that letter gets in [loved one's] hand\ldots how many people have already looked at it and sorted through it? I don’t really like that feeling.''\textit{---F2}
\end{displayquote}

\noindent 
However, system-impacted people have no option but to continue utilizing and depending on the digital devices that are available. P7 noted that in prison:

\begin{displayquote}
``You give consent by being there. Nobody asks you for any kind of consent. Do you want a phone call? Do you want an email? Do you want music? Consent is implied\ldots if you want any of that.''
\end{displayquote}

\noindent P8 expressed that ``When it's the only option, what else do I do?'' We discuss how this lack of free choice \textit{if} they want to communicate with their loved ones, and the lack of free choice over \textit{what technology} and provider they choose to use, leads to a sense of apathy and reluctant acceptance in Section \ref{ssc:apathy}.

\subsubsection{Technology Changes Increase Surveillance}\label{ssc:flux} 
All participants mentioned initially using the physical mail system and wall-mounted telephones to communicate with their loved ones outside the prison. This changed over time; many described the introduction of JPay devices (Figure~\ref{fig:JPAY_tablet}) as a means to communicate with loved ones.

Following a 2015 buyout of JPay by Securus~\cite{jpayAquisition2015}, many participants' JPay devices were replaced with Securus tablets. Updates and changes in technology did not necessarily lead to the improvements participants hoped for. For example, the changeover from JPay to Securus after the buyout led to a decline in quality and usability of the technology:
\begin{displayquote}
``Once it switched over to Securus---can’t believe we're gonna say this---that we actually miss JPay. Securus just seems like it has, they both have their own set of problems, but it just seems like the technology has not gotten better, it's gotten worse, with Securus.'' \textit{---F2}
\end{displayquote}

\noindent The changeovers negatively affected  usability and increased surveillance and censorship, leading to more stringent restrictions for participants. P6, for instance, described how communication had become more restrictive:
\begin{displayquote}
``Nowadays, it being Securus and the changes that they've made, people on the outside are not allowed to initiate contact\ldots You used to be able to send emails out; you could talk to them if you knew their email address; there's no option for that anymore.'' 
\end{displayquote}

\noindent The perception of increasing censorship was also felt during new video and audio communications. F2 described: ``With JPay, [it's] more censored, even more so.'' Despite the outside world advancing technologically, our participants felt their technology was regressing; P9 described how ``it was really fascinating to watch backward technology while [time] was going forward.''

\subsubsection{Surveillance Leads to Communication Delays}\label{ssc:seamless}

Because messages are surveilled for content, participants experienced long delays in their communication being sent and received, if they are received at all. Moreover, participants lacked information about whether their messages were censored, and how. As P1 described: 
\begin{displayquote}
``Even if you message on JPay but you said the wrong thing on there\ldots Your messages may come through slower\ldots or they won't come through at all. Sometimes, they don't even let [you] know that.'' 
\end{displayquote}

The delays and drops in message exchange due to censorship belie the idea of seamless communication through technology according to some participants. As P8 reflected:
\begin{displayquote}
``\ldots you think, `Oh, I can write this email' and\ldots send it off into the ether; it could sit in the mailroom for days if it had something in it that alerted [the prison officials]. And then you wouldn't even know that your email didn't go out, let alone why you weren't getting a response, or if it did go out, but now they're watching you.'' \textit{---P8}
\end{displayquote}

\noindent We elaborate more on the working of censorship through technology surveillance in Section \ref{subsec:adverse}.

\subsubsection{Other Consumer Protection and Autonomy Issues: Usability Challenges and Financial Burdens}\label{ssc:userissues} 

The monopolies of few companies on the market for digital devices in U.S. prisons, combined with the technology access restrictions that stem (at least apparently) in part from prison security concerns, also led to serious usability and financial challenges. Both were such pervasive and taxing issues that all of our participants brought them up without being prompted.

In terms of hardware usability, the tablets were prone to easy breakage. P5 described that: ``one of the worst things that would happen would be the battery swelling\ldots it'd swell up and crack the screen.'' 
The constant hardware problems forced participants and families to pay for devices anew, which, as P4 implies, benefits Securus: ``Tablets would break because they're poor quality. They always broke because then we would have to re-buy them.'' 

Participants further experienced software issues whereby their applications often froze, their purchased data was lost, or telecommunication features worked poorly. F2, a family member, described: ``we deal with this almost every day---sometimes, he'll call, and it will be dead air. Or he'll have to call back. And I don't realize. On his end, he's probably trying $20$ or $50$ times to try to get through.''

Finally, the physical implementations of devices caused discomfort. F1 described the physical contortions that their incarcerated spouse had to undergo when video calling at a poorly designed kiosk with a payphone handset attached to a cord that is only $12$ inches long: ``We typically won't even go like the full $30$ minutes at the call because his neck, like you can see how uncomfortable it is, holding that.''

All participants, without being prompted, also mentioned the costly pricing and hidden fees when using prison technologies. 
This was particularly worrisome for participants given that, generally in the non-prison environment, the same services (such as email) are free. P4 described how these exorbitant costs compared to their low wages, observing:

\begin{displayquote}
 ``[Sending a message costs] 50 cents, and if you do that everyday, then of course that's like 15 dollars a month, and in a place where they pay you, I think it's 18 cents an hour, 21 cents an hour ... So I mean those things balance poorly.''   
\end{displayquote} 

P11 similarly expressed that ``They cost more than what we are paid. For no reason at all.''
The costs and fees also burdened the loved ones of incarcerated people, who often used these technologies to send money. P8 explained that: 

\begin{displayquote}
``When I was first incarcerated, if you got money sent in, it was $35$\% taken out, and it changed to $55$\% within a couple years of my incarceration. Which means if you get 100 bucks sent in, you're only getting $45$ out of it 
\ldots it's really hard to justify that to a family\ldots already struggling.''
\end{displayquote}

P3 summarized the sense of unfairness as well when asking: ``How are you going to offer something to people and charge them when you know that they don't have money?''

As such, participants were financially burdened by both the fees charged by Securus, and the taxation policies of DOC. President Biden attempted to put a cap on these fees \cite{MarthaWrightReed}, although President Trump reversed this in 2025 \cite{motherjonesTrumpsScraps}.

\subsection{Prison as the ``Man'' in the Middle}
\label{subsec:rq2}\label{subsec:surveillance}

Before Bob receives Alice’s message, it is intercepted by Department of Corrections (DOC) staff, i.e., the “man in the middle” who exerts technical and bureaucratic control over the communication channel. Participants perceive this interception as a symptom of power asymmetry wherein power is unilaterally exercised by staff~(Section \ref{ssc:power}); their perspective was that that staff often operate from misinformed or overly punitive threat models
(Section \ref{ssc:staff}).

\subsubsection{Power in the Hands of Officers}\label{ssc:power}
The extent to which participants can access technology, and how they are surveilled and censored, depends on the discretion of the staff and officers overseeing them. P4 observed that: ``The power dynamic that exists within the carceral circumstance is one that entirely leaves it up to the whim of whatever staff member it is.'' This power dynamic fosters feelings of scarcity and competition around the technologies. P1 described this in the context of the kiosks, saying: ``Oftentimes, [the kiosks] were so overloaded that one of them would be down.'' They noted that this scarcity led to altercations and competition: ``There were so many altercations, so many fights, so many assaults, just because of that.'' 

The increase in altercations as a result of device scarcity, rule-setting, and restriction can be seen as ironic given that the justification for these restrictions is often ``safety'' and ``security.'' P1 hypothesized that the scarcity was engineered since officers profited from the chaos, underscoring the lack of trust between incarcerated people and prison staff:

\begin{displayquote}
 ``They [the Department of Corrections officers] run up the fights, they run up the infractions, and they say, `Oh well, we need this much more in the budget because we need to hire this many more people for security or the guards need to work overtime.' '' 
\end{displayquote}

\noindent These power dynamics apply to censorship as well. Officials have discretion about whom and what they censor, oftentimes perceived as wielding this power with no impartiality, as P1 explained: ``There’s one individual who for some weird reason, may just dislike you. And that’s the thing about it, there is no recourse\ldots So yeah, so they are more likely\ldots to listen to our calls.''

Thus, participants felt that technology was often weaponized against them to restrict their privacy and augment the power of the officers. They observed that securing against digital threats seemed more important to prison staff than incarcerated people having the ability to communicate with their loved ones, lawyers, or community members. P2 summed this up by observing that ``the system will choose security over humanity every time.''

\subsubsection{Censorship Feels Arbitrary}\label{ssc:arbitrary} The power of individual officials to control levels of surveillance and censorship was combined with a sense of arbitrary policies; the decision regarding who was censored and how seemed arbitrary to participants. For example, P4, in a focus group with P5, said ``[P5] and I can send the exact same image to two different people, and one will get it on the exact same day, and the other will get it rejected.'' 

F1 also recounted a time they tried to communicate with their incarcerated spouse, and the message was censored: 

\begin{displayquote}
``I mean this was like a very basic email that I had sent my husband, like nothing suggestive. Like, `Call me at whatever time' or something like that. And [officers] sent [me] this thing that there was like, obscene whatever, or something, I don't know, it was just nuts.''  \textit{---F1}  
\end{displayquote}

\noindent Such seemingly arbitrary decisions contribute to further power disparities, allowing officers to control what people can see and when. They can, with little warning or recourse, restrict communication with loved ones or legal counsel. For instance, P4 explained that: 

\begin{displayquote}
``[My spouse] would send legal work that I was working on; she'd take pictures of it so that I could edit it. [A]nd there came a point where [an officer said] said, `Yeah, we're rejecting this, it's too much text to review.' '' \textit{---P4}
\end{displayquote}

\subsubsection{Prison Staff May Hold Inaccurate Threat Models} \label{ssc:staff} 
Staff and officers wield power and use technology as another means to exert it: they surveil, enforce censorship, and arbitrarily revoke access. Participants expressed that prison staff's misconceptions about technology may exacerbate these issues. P8, for example, explained that they did not think prison officials had a factual understanding of the technology, and, because of this, took away access to it, saying:  ``They slowly were like `Okay, wait a second, we think something tech-y is happening here, but we don't know what. So we're gonna take away all these things.' ''

P8 continued by explaining how the officers are taught to restrict the digital rights and access of incarcerated people due to security fears while actively violating the security and privacy of incarcerated people: 

\begin{displayquote}
``So the officers actually have all the access to everything that could hurt us.  And yet they are taught to be afraid of us. A lot of it is around your personal information, which to me links directly to tech.\ldots They're literally thinking, `You might get my login for something; you might see my password; you might see the information that's in my employee account'\ldots They're not thinking of you getting a hold of somebody's piece of mail; they're literally thinking of their digital identity.'' \textit{---P8}
\end{displayquote}

\noindent Finally, technology is perceived as both overwhelmingly powerful and yet simultaneously fragile and easily compromised by any unsupervised incarcerated person. This lack of understanding may contribute to staff responding with fear:

\begin{displayquote}
``There were two staff members\ldots talking about moving everything to Wi-Fi. And they said, `Well gosh, now they're gonna be getting onto the Internet, they're gonna be hacking into the Internet and doing a bunch of identity theft'\ldots like they're making this whole scenario about things that inmates are going to be doing on a secure platform\ldots the staff's ignorance about the technology is a counter to the benefit of it.'' \textit{---P5}
\end{displayquote}

\subsection{Downstream Impacts}
\label{subsec:adverse}\label{subsec:rq3}

The final component of the Alice and Bob model (Figure 2c) shows the subsequent effects of the staff's decisions to censor or block Alice’s message. These decisions, described by participants as often arbitrary and unpredictable, have profound downstream consequences that threaten core aspects of personal well-being: engendering self-doubt (Section \ref{ssc:selfcensor}), eroding community ties (Section \ref{ssc:commthreat}), hindering rehabilitation (Section \ref{ssc:rehabthreat}), and feelings of resignation and powerlessness in the face of technology access (Section \ref{ssc:apathy}). 

\subsubsection{Internalized Threats of Self}\label{ssc:selfcensor} 
Participants shared that the system of censorship within prisons conveys a clear message: the use of technology by incarcerated people is inherently dangerous and untrustworthy, necessitating surveillance. This narrative is further reinforced when officers can revoke technology access under the guise of mitigating security risks. P8 described how: 

\begin{displayquote}

``I sent an email, but that email didn't actually go somewhere; it goes to the mailroom. It goes to somebody else's inbox, for them to look at it first. And for them to determine if they think it's okay to be sent out. And so you don't even have that kind of instant gratification of like, okay I sent that thing. You don't even know if it goes, and that breeds that distrust. Both of self, because it's like, okay, did I do that, right? And then of the tech, because it's, `Okay, is the tech malfunctioning?', and how will I know if it's doing the right thing?'' \textit{---P8}     
\end{displayquote}

\noindent These fears became so internalized that they remained even after P8's release: 

\begin{displayquote}
``So when I was getting on the Internet and searching things when I first got out [of prison], I was like `Oh, should I do this?' Like I just was instantly thinking, I'm self-monitoring. And I was really afraid to get on social media because what if I'm not allowed on here?'' \textit{---P8}   
\end{displayquote}     

\noindent Other participants reflected on similar trends of self-censorship and how the incessant psychological double-think lingers. As P2 described:
\begin{displayquote}
``Landing here in my reintegration journey, there has been [psychological] overrides due to developmentally being arrested through incarceration. There’s\ldots an override to show up, speak up, be seen, do these things, and this is part of the threat of us being censored from just being human.''\textit{---P2}
\end{displayquote}

\noindent P2 likened the ``psychological trauma that’s being taken out on our bodies'' to that of ``treading water'' and ``drowning.'' Similarly, in P8's words, censorship created ``the panopticon\ldots at its finest.''

\subsubsection{Threats to Community}\label{ssc:commthreat}

Surveillance and censorship of communication technologies inhibits participants from fostering their intimate, human relationships.

P2 expressed that:
\begin{displayquote}
``Yeah, I would just highlight the censorship of like, I was married while I was inside. There’s a whole process to even get correspondence with my wife. And then, to speak about intimate things, our mail was rejected. And it’s like, I’m human, and I have a wife, and they stopped that.'' \textit{---P2}
\end{displayquote}

 According to P10, the prevailing expectation from Securus and DOC is that:  ``10 minutes is enough to build a relationship, right? But it’s not'' (P10). The consequences if they chose not to use this technology, or if their access was restricted or revoked, was grave: ``My body was being affected by how much anxiety and fear [I had] and [I was] so scared that my boys weren’t gonna remember me'' (P10).

 Surveillance and censorship further prohibited those seeking outside legal council or support. P5 voiced  described how censorship prevented him from seeking the legal council he required, observing that his friend sent him ``a picture from the\ldots attorney general's website of this year's people who were on the clemency list for 2024. They [prison officials] rejected it as third-party information.''

\subsubsection{Threats to Rehabilitation}\label{ssc:rehabthreat} 
Participants expressed that the exposure to surveillance and censorship, in addition to arbitrary technology restrictions, leads to a lack of digital literacy within and beyond prisons. This does not adequately prepare them for the technological world once released~\cite{ogbonnaya-ogburu_nonprofits, ogbonnaya2018returning, ogbonnaya2019towards}. 

This digital disadvantage is partially due to the loss of privacy and agency that system-impacted people experience since their life stories and their pasts are forever documented online, accessible to the public (echoing~\cite{gurusamiWeb}). The fear of being discovered as formerly incarcerated relates to participants experiencing difficulties in using technologies once released. As P8 recounts, the lack of digital literacy created a feeling of ``being outed'' or ``othered,'' stating: ``That was my reaction to almost everything: They're gonna know that I was in prison because I don't know how to do this.''

Participants also perceived that the surveillance and censorship continued, even post-release, to undermine  efforts for peaceful reintegration. Echoing Owens et al.'s~\cite{owens22usenixsecurity} findings on post-release electronic monitoring, P8 discussed the anxiety induced by post-release electronic monitoring applications:

\begin{displayquote}
``I have to submit an entire schedule, and I'm beholden to it. Where it will take me 10 minutes to get from here to here, and then 15 minutes to find parking, and then 30 minutes at a meeting, and then 45 minutes to get back home. 
\ldots and that is another way of surveilling, where that gets loaded into the computer\ldots So all these invisible ways, again, that technology, we're told that it's for the betterment of our lives and that this will make things easier. And yet when we get surveilled by it, it's used in this really punitive way.'' \textit{---P8}
\end{displayquote}

\noindent The threat of surveillance and censorship leads participants to fear and second-guess their actions. It prevents them from adapting to technologies, since they perceive them  as means to continue the cycle of surveillance and censorship experienced in prison. P8, for instance, explains how easy it was to be taken advantage of on social media since they had no prior experience with and were afraid of technology: 

\begin{displayquote}
``I was really afraid to get on social media because what if I’m not allowed on here? What if I run into someone who knows me?\ldots I ended up getting trolled, really bad actually, on Facebook. So I don’t use Facebook anymore\ldots One, I feel like I shouldn’t be doing this, and then it’s reinforced because the reaction I end up getting [is] like, `Oh, you did it wrong'.'' \textit{---P8} 
\end{displayquote} 

\noindent This lack of access and fear of technology can thus deepen the digital divide for system-impacted people and make them more vulnerable to social isolation or even online scams.

\subsubsection{Apathy and Resignation}\label{ssc:apathy} 
The reluctant acceptance of technologies stems from a feeling that it is unlikely that problems will be adequately fixed. Participants felt powerless in the face of dismissal and against the persistent usability issues and appeals of censorship that offered them no resolution. P4, for instance, described the process they went through to address an issue, noting that: ``You find yourself back in this place where you're completely defeated, you've found no recourse, nobody is accountable, it's never anybody's fault.'' P10 was told their were unable to appeal the decision to revoke their communication with their family: ``I sent complaints and everything. Then I found out that I couldn’t even appeal it.''  P2 also felt that complaining or raising injustices inside the prison only left them feeling depressed, articulating that, for the incarcerated person: ``If he stops too long to gripe\ldots then he drowns.'' 

This feeling of powerlessness paradoxically left participants feeling grateful and appreciative for the same technologies that harm them, a phenomenon that P5 dubbed ``technology Stockholm syndrome.'' P4 explained:

\begin{displayquote}
 ``So, all of these hurdles, all of these things just to say if a phone call works well for a full 20 minutes, you just have to kinda consider yourself lucky\ldots like you shouldn't be grateful that somebody didn't kick you after they knocked you down.''  
\end{displayquote} 

\noindent With meager expectations, the smallest successes made them feel lucky to have access to anything at all. F1 elaborated, explaining that: ``A lot of times I'll be like, `Just be glad for what you get.''' P7 further contextualized how access to technology within prisons is often viewed as a substitute for educational opportunities. Additionally, they connected this access to broader issues of consumerism and technology addiction in the outside world: 

\begin{quote}
``I don’t know that [technology access] prepared anyone specifically, it probably prepared them to be good consumers. And how to get lost in your phone for hours. Now they have books and movies, and the movies can run up to 25 bucks\ldots or something like that, and they're shitty from what I hear. I don’t know that it prepared me [for release] in the same way a school would.'' 
\end{quote}

\section{Discussion}\label{sec:discussion}

In general, we found that the devices currently deployed in prisons (in our geographic area) are used as tools for surveillance and censorship, which ultimately impede community building and rehabilitation efforts for system-impacted people. Further, the technology is replaced or rescinded for opaque reasons and without notice for its end users due in part to technology misperceptions held by the staff. These justifications often revolved around ``threats to [prison] security.''

We now discuss two tensions that surfaced during our investigation, and recommendations for their mitigation, aimed at the research community, policymakers, prison officials, and prison technology vendors. 

Although we make recommendations, we believe that no matter how usable, accessible, and modern the technology is inside of prisons, the carceral system is an inherently traumatizing force in U.S. society. We echo Owens et al.~\cite{kentrellCHI} vis-a-vis recommendations: ``While we can offer design and policy recommendations to make communication easier and more privacy-preserving for [family members of incarcerated people] and their incarcerated relatives, the most effective solution is one that reduces the number of people who are subjected to carceral surveillance in the first place.'' 

\begin{subsection}{Observed Tension: Harms from No Technology versus Harms from Technology} 
\end{subsection}

As explored extensively in prior work~\cite{reisdorf2022digital, ogbonnaya2019towards, seo2022returning, ogbonnaya-ogburu_nonprofits, reisdorf2018digital, grierson2022design, anuyah2023characterizing}, system-impacted people face a significant uphill battle when adapting to a rapidly changing digital world. This is especially true if they have been incarcerated over the last three decades, as digital devices became necessary for navigating society. Many of our formerly incarcerated participants mentioned difficulties post-release interfacing with technologies: from creating an e-mail address, to using QR codes at restaurants, to procuring housing or establishing bank accounts. Prior work explores how the lack of access to technologies for incarcerated people is deepening the digital divide~\cite{ogbonnaya2019towards, seo2022returning, reisdorf2018digital, grierson2022design}. Some recent papers~\cite{gautam2024enhancing, nisser2024prisons, martinez2024engaging} and non-profits~\cite{ameelio, UnlockedLabs} aim to close this digital divide by co-designing and deploying digital tools for incarcerated people. 

We extend this prior work to demonstrate that, while harms from a lack of access to technology exist, other harms from having exposure to technology emerge. Digital devices that system-impacted people can access lead them to experience surveillance and censorship, strain their community ties, and further threaten their rehabilitation. According to study participants, the tools become sites of power contestation and arbitrary and punitive bargaining chips by officers.

In sum, if a technology is introduced into prisons to support incarcerated people and close the digital divide, \textit{we recommend it be accompanied by considerations of its socio-technical context, and in particular, how the (even well-intentioned) technology could be used in harmful ways.}

\paragraph{Recommendation: Free, open, digital and non-digital opportunities.}
As P7 expressed in Section \ref{ssc:apathy}, access to technology encouraged consumerism rather than providing meaningful preparation for release. They underlined that a formal education, like a school, would have better prepared them.
Technology access was not a reprieve from consumerism, financial stress, coercive choice, or a replacement for real educational opportunity. We therefore recommend that \textit{digital opportunities should be provided in addition to non-digital educational opportunities}. \textit{Information should also be made available about the harms of technology,} from technology addiction and consumerism to privacy vulnerabilities, scams, phishing, and the like. 
Moreover, after release, systems-impacted people remain under surveillance of their social media, location, browsing activity, and community ties~\cite{owens22usenixsecurity, socialmediasupervision, Fussell_2020, communitysupervision, randcommsup, docsocialmedia,Justfutureproject}. Developing a stronger awareness of their security and privacy vulnerabilities can help them navigate these risks and support successful reentry.

We also believe \textit{educational opportunities, notably threat modeling, could be beneficial for prison staff}. In Section \ref{ssc:staff}, we note how participants often felt that prison officials held misconceptions about their potential use of technology. Yet, as noted in Section \ref{ssc:power}, staff are the ones responsible for enforcing its use. We believe education for prison officials could be useful to explore the capabilities and limitations of certain technologies, including dispelling common misconceptions about the misuse of technology inside prisons. Indeed, some prior work has shown that educating staff about technology can foster greater support and reduce the stigma for incarcerated people's use of technology~\cite{davies2017implementation, knight2024digitalising, antojado2024future, unicri}.

Regarding their pricing, prior work also highlighted the financial burden that available technologies place on already struggling system-impacted families~\cite{devuono2015pays, sobol2018connecting, cortina2022want}. This is particularly concerning given that poverty is a predictor of incarceration, and also increasingly an outcome, in part due to such exploitative pricing schemes~\cite{prisonpolicyinitiative,devuono2015pays}. 

We believe that making this technology free and/or affordable is not an unrealistic expectation: the San Francisco Public Library recently adopted a new no-hidden-costs, completely free tablet program for all locally incarcerated people~\cite{SF_Public_Library_2023}; in September 2022, California joined Connecticut as the second state to make phone calls free in state prisons~\cite{Family_Friends_Services_2024}; and in July 2024 the FCC approved new caps on prison phone call prices and banned some kickback schemes in response to the Martha Wright-Reed Just and Reasonable Communications Act of 2022~\cite{2024FCC, MarthaWrightReed}, although President Trump revoked these bans/caps in 2025 \cite{motherjonesTrumpsScraps}.

\vspace{0.15in}

\subsection{Observed Tension: The Customer Is Not the End-User}

The interviews we conducted suggested that prison policies and regulations appeared to be subject to the whims of officers (Section \ref{ssc:power}) and that interviewees felt there was no accountability for or recourse to change usability problems and censorship (Section \ref{ssc:apathy}). 

Under the current incentive structure, prison officials enter into contracts with technology companies, making the companies accountable to the officials and staff, not to actual system end-users; end-users are the incarcerated people and their families, whom Owens et al.\ describe as ``both literally and figuratively a captive audience''~\cite{kentrellCHI}. Our findings reveal how this arrangement intensifies censorship and financial burdens (e.g., having to pay to resend a message after it is censored), benefiting both the prison officials and the contracted technology companies but not end-users. 

\paragraph{Recommendation: Transparency and accountability to the actual end-users.} 
We previously noted how users are often not notified about whether their messages have been censored (Section \ref{ssc:seamless}). Though transparency regarding block-lists of words that trigger censorship could help reduce the financial burdens faced by incarcerated people (from not having to pay to resend messages), we recognize that this approach may not be desirable for prison staff due to concerns about circumventing censorship. Nevertheless, we believe that \textit{incarcerated people should be told whether and why their messages were censored.} We also recommend \textit{establishing clear guidelines for disallowed language} to ensure that prison officials' decisions are not arbitrary. Furthermore, we call for more research to verify the claims of prison officials that censorship prevents illegal or otherwise harmful activity; \textit{more research is needed to understand if the censorship accomplishes its claimed benefits, and does not just prevent system-impacted people from openly speaking with their loved ones and lawyers. }

Without openness and transparency, security and privacy research has long argued that ``security through obscurity'' is a misguided strategy~\cite{whycryptosystemsfail, neumannmercuri, Kerckhoffs_1883, katz2007introduction, hall2025pitfalls}. In that vein, we recommend that not only the aforementioned censorship guidelines, but also \textit{Privacy Impact Assessments (PIAs)} --- i.e., assessments that every U.S. federal agency must fill out regarding personal data collection and privacy --- \textit{be required for prisons and their contractors and made publicly available}~\cite{hhsPrivacyImpact}. We also recommend that these assessments be extended beyond the scope of personal privacy, towards \textit{including all potential harms and risks the technology may impose.} This could include the financial and social harms described in Section~\ref{subsec:adverse}. 

\textit{We further recommend that prison contracts with private companies be subject to a mandatory public comment period before a private company's bid is chosen.} In particular, these contracts should incorporate input from system-impacted community members and be awarded in part based on robust research about the potential harms of the contracted technology to its end-users. 


\paragraph{Recommendation:  Consider and audit more carefully the use of ``security'' and ``safety'' for the carceral institution as a justification to violate the security, privacy, and safety incarcerated people.}
Finally, we noted the repeated reference to 
``security'' and ``safety'' threats \textit{to prisons} when asking participants why they believed their own security, privacy, and safety were violated. These violations ranged from officials searching their belongings and messages, intense surveillance and restriction on device use, as well as coercion, competition from manufactured scarcity, and lack of autonomy over their personal information and documents. However, the underlying prison security concerns may not be founded in actual data about real threats, and moreover the implied threat model creates opportunities and excuses for abuses of power. 

If prisons intend to reduce tension and violence inside their walls, we worry that such intense restriction, power contestation, and competition over the very technology that connects people to their support systems is not helpful. Such frustration, isolation, lack of autonomy, and scarcity could, in fact, be detrimental to the prison's stated goals of maintaining a peaceful environment, as we discussed in Section \ref{ssc:power}.

In sum, we believe that work is needed to address the privacy, security, usability, and financial issues stemming from technology available to incarcerated people, and we believe that this work can take many achievable forms: including educational reforms, consumer protection safeguards, policy implementations, technical solutions, bug fixes, and more.

\section{Conclusion}
To explore the security, privacy, and associated impacts of technology inside the carceral system, we conducted focus groups and interviews with 17 formerly incarcerated people and relatives. Our findings show that participants experience pervasive forms of surveillance and censorship related to the available technologies. They describe shifting and arbitrary technology censorship policies that strain their relationships with their communities and threaten their future rehabilitation; 
these challenges exacerbated feelings of dehumanization, self-doubt, and mistrust towards technology. We end by recommending ways to modify these technologies and policies to better serve incarcerated people and their loved ones, and thereby ultimately for society as a whole.

\section{Acknowledgments}
Thank you to our participants. Thank you to the members of the Security and Privacy lab at UW. Thank you to the currently incarcerated people at the Monroe Correctional Complex. This work was done as part of the Center for Privacy and Security for Marginalized and Vulnerable Populations (PRISM), supported by the National Science Foundation under Frontiers (award number 2205171).

\bibliographystyle{plain}
\bibliography{ref}

\appendix

\section{Interview Materials}
\label{appendix:protocol}

\subsection{Consent \& Disclaimers}
In this study, our goal is to understand the perception that users of communication technologies within prisons have around usability, security and privacy. Throughout this interview, we will ask you questions to understand your experiences with these technologies. Before we get started, we want to mention that:

\begin{enumerate}
 \item The interview will be around one hour.
 
 \item With your consent, this interview will be recorded and then transcribed. 

 \item We can delete things from the transcript that you’re not comfortable with being included.
 
 \item The interviews will be transcribed fully manually; we will not use any third-party transcription services. 
 
 \item To protect your anonymity, we will use alias’ for all identities. Please let us know if you have an alias in mind you’d prefer to use for yourself. We will also not store names or email addresses together with the recordings.
 
 \item We are also not going to share the transcripts and recordings with anyone else beyond the research team.

 \item We can share our contact information with you if you want to follow-up (or share the contact information of the advisor), including our alias’ on the Signal encrypted messaging app, if you prefer that over texting or emailing.

 \item We will also compensate you for the time you spend with us. The compensation is set at \$35 in the form of a Visa Gift Card. These gift cards can be used like any debit or credit card both in stores and online. 
\end{enumerate}

All that being said, do you consent to be interviewed? Do you consent to be recorded? Do you consent us to transcribe these interviews? You can always rescind your consent at any point in the process and still be eligible for the compensation. 

We also want you to be as involved in this project as you want to be---we will send you the manuscript we write via email before we send it to publication to make sure it accurately reflects your experiences. You will have two weeks to submit your feedback. 

We recognize that we are benefiting from this project for our research, and we want to make clear that our goal is to raise awareness to the experiences of incarcerated people with this technology, and bring this knowledge to new audiences. We also know that researchers, especially when working with marginalized groups, have a history of parachuting in and then leaving. We want to make clear that anything we can do for you is paramount to us, and we hope to build a long-lasting relationship with you all.  

If, throughout the interviews, you need to take a break, you want to stop early, grab some water or something to eat, you can definitely have a time-out. We have water, snacks, and tissues available for you. 

If anyone has questions, thoughts, comments, or any discomfort whatsoever, please let us know and feel safe to do so! 

\subsection{Interview Protocol for Formerly Incarcerated Participants}

\subsubsection{Available Technologies}
\begin{enumerate}[label=\intq{\emph{Q}}{\arabic*}.]
    \item Where were you incarcerated? Did that change over time?
    \item What technologies did you have available to you? Did that change over time or based on the place you were incarcerated at? If yes, what caused the change? Did the technology have varying security levels?
    \begin{enumerate}
        \item What did you use the tablets for generally? Was there anything you wished you could have used the tablets for but couldn't? Were there restrictions on using the tablets or what you could use them for? 
        \item In order to use the tablet, did you have to do something specific, like enter a password or show it your face? How does account management work for these apps and technologies? Who creates/knows the passwords? Do you have things like 2FA? Do multiple people share the same device? Or are they assigned 1-1?
        \item Were there certain times you could use the technology, or other restrictions? Was it like a regular schedule, access whenever they wanted, access only under certain circumstances, etc?
        \item What other apps were on the tablets?
        \item Did the tablets have cameras or microphones? Were those available to use? Did they have permissions?  Did you ever feel some type of way about those devices?
    \end{enumerate}

    \item Was there anyone on the outside that you communicated with? If so, what technology did you use to communicate with them? Did that change over time? Why and how did you choose some over others?
    \begin{enumerate}
        \item To what extent and for what purposes do you use Getting Out / GTL and JPay / Securus? How frequently do you use the apps? Did that change over time?
        \item Were there any other options if you didn't want to use Securus etc?
    \end{enumerate}
    \item Was there anyone on the inside that you communicated with? If so, what technology did you use to communicate with them? Did that change over time? Why and how did you choose some over others?
    \item How did using those apps change for you when you were inside to when you were released? Do you still use the apps now to talk to friends inside? How does the experience differ on the tablet inside to your smartphone outside?
    \begin{enumerate}
        \item What was the process like while you were inside? Can you take me through from \textit{Step 1: I want to send a message} to \textit{Step n: This message was received}? This description should also include how account management works, e.g., who creates/knows the passwords? Is there a 2FA?
        \item What is the process like now on the outside if you want to communicate with someone inside?Can you take me through from \textit{Step 1: I want to send a message} to \textit{Step n: This message was received}. This description should also include how account management works, e.g., who creates/knows the passwords? Is there a 2FA?
        \item How have your perceptions of the app changed over time, especially pre-release versus post-release?
    \end{enumerate}
\end{enumerate}

\subsubsection{Perceptions of Technologies}
I am now going to ask you about how you felt about these apps. I know some people may have had good experiences with these apps. Other people may have had bad experiences with these apps. Some people may have had both good and bad experiences. I am interested in the experiences you had, either good, or bad, or both.
\begin{enumerate}[label=\intq{\emph{Q}}{\arabic*}.]
    \item How do you feel generally about these apps and/or the other technology available to you?
    \begin{enumerate}
        \item What is a specific bad experience you've had with the app and/or other technology?
        \item What is a specific good experience you’ve had with the app and/or other technology?
        \item What did you like and dislike about the technology you used to communicate?
    \end{enumerate}

    \item What types of things did you talk about with these technologies? Are there topics of conversation you wanted to discuss but decided not to with this technology?
    \begin{enumerate}
        \item When you wanted to communicate with someone, and had the technology available, did you ever consider \textit{not} using the technology to communicate with that someone, and why?
    \end{enumerate}
    \item How did it feel to use this technology? Frustrating? Encouraging? Secure?
    \begin{enumerate}
        \item Did you ever feel uncomfortable using the technology? Why or why not?
        \item What strategies did you use (if any) to manage your concerns? 
        \item Did you ever observe someone else struggling to use the technology? Did you ever show them something on it?
    \end{enumerate}
    \item What kind of information do you think these apps are collecting? Who do you think has access to the data that is being collected?
    \begin{enumerate}
        \item Did any prison staff ever say something to you that made you think that maybe they read your messages?
    \end{enumerate}
    \item What is the money / payment process like and what do you think of it? How much did it cost? Did the cost impact your communication?
    \begin{enumerate}
        \item How were the rules about what was allowed/what you’d be charged for communicated to you?
    \end{enumerate}
    \item If you were asked to describe this technology to a friend, who has never used the technology and has never been incarcerated, how would you describe it?
    \item How did technology impact your life before you were incarcerated, while incarcerated, and after?
    \item How did your engagement with technology change during the COVID lockdown and pandemic?
    \item If you could design a communication technology for incarcerated people to use, what would it be like?
    \item What else do you want us to know?
\end{enumerate}

\subsection{Interview Protocol for Loved Ones of Incarcerated Individuals}
\begin{enumerate}[label=\intq{\emph{Q}}{\arabic*}.]
    \item Where and when (start and end dates) is/was your loved one incarcerated? Did that change over time? 
    \item What technology do/did you use to communicate with your loved ones on the inside? Did that change over time? Why and how did you choose some over others?
    \begin{enumerate}
        \item To what extent and for what purposes do you use Getting Out / GTL and JPay / Securus? How frequently do you use the apps?
        \item Were there any other options if you didn't want to use Getting Out or JPay?
    \end{enumerate}
    \item What has your general experience with these apps been?
    \begin{enumerate}
        \item What did you like and dislike about the technology you used to communicate?
        \item How did it feel to use this technology? Frustrating? Encouraging? Secure?
        \item What is the communication process like? Can you take me through from \textit{Step 1: I want to send a message} to \textit{Step n: This message was sent}?
        \item Are there topics of conversation you didn’t use the technology for?
    \end{enumerate}
    \item What kind of information do you think these apps are collecting? Who do you think has access to the data that is being collected?
    \item What is the money / payment process like and what do you think of it?
    \item How did this all change during the COVID lockdown and pandemic? 
    \item If you could design a communication technology for incarcerated people to use, what would it be like?
    \item What else do you want us to know? 
\end{enumerate}

\section{Qualitative Codebook}
\label{appendix:codebook}

See Table \ref{tab:codebook}.

\begin{table*}[tb]
\centering
\footnotesize
    \begin{tabular}{@{} llp{9.7cm} @{}}
        \toprule
        \textbf{Top-Level Code} & \textbf{Code} & \textbf{Definition} \\ \midrule
        Communication Tech. & 
        
        Mail & Procedure/practice related to physical items or documents sent through the post. \\ \cmidrule(l){2-3} & 
        
        CorrLinks & Procedure/practice related to the usage of the Corrlinks system. \\ \cmidrule(l){2-3} & 
        
        Messaging & Procedure/practice related to messaging apps such as Securus or JPay (not Corrlinks). \\ \cmidrule(l){2-3} &
        
        Kiosk & Procedure/practice related to utilizing the kiosks for communication purposes. \\ \midrule

        Media \& Education & 
        
        Laptops  & Procedure/practice related to utilizing laptops for education or as a form of media. \\ \cmidrule(l){2-3} & 
        
        Legal/policy library & Procedure/practice related to legal/policy library applications for education or as a form of media. \\ \cmidrule(l){2-3} & 
        
        eBooks & Procedure/practice related to eBooks on tablets for education or as a form of media. \\ \cmidrule(l){2-3} &
        
        Television & Procedure/practice related to utilizing television for the purposes of education  or as a form of media. \\ \cmidrule(l){2-3} &

        Music & Procedure/practice related to music for education or as a form of media. \\ \midrule

        Benefits & 
        
        Faster communication  & Technology that reduces the time between sending/receiving of information compared to conventional or previously used methods. \\ \cmidrule(l){2-3} & 
        
        Training for re-entry & Technology that helps teach skills that would then be transferrable to the ``real world'' when individual is released. \\ \cmidrule(l){2-3} & 
        
        Cultural touchstone & Technology that allows individuals to help connect to cultural moments from the ``real world'', such as watching popular movies. \\ \midrule

        Harms & 
        
        Usability problems  & Issues or deficiencies in the design of the technology that hinder its ease of use, efficiency, and effectiveness (e.g., tablets overheating when charging). \\ \cmidrule(l){2-3} & 

        Censorship  & Examining mail, messages, etc. and confiscating items, suppressing messages, etc. \\ \cmidrule(l){2-3} & 

        Scarcity  & Competitive behaviors that arise a limitation or scarcity of the technology with the potential to escalate into conflict. \\ \cmidrule(l){2-3} & 

        Expensive  & Technology requiring significant financial resources. \\ \cmidrule(l){2-3} & 

        Bargaining chip & Technology utilized within the hierarchical structure of the prison, especially by guards, to negotiate, influence, or assert dominance over those incarcerated. \\ \cmidrule(l){2-3} & 

        Adaptation & Introduction of new technologies that necessitate incarcerated individuals to learn both technically how they work, as well as adjust to new use-related practices. \\ \cmidrule(l){2-3} & 

        Insufficient knowledge  & Lacking sufficient information to understand how and for what technology might be used. This applies to both incarcerated individuals and prison staff. \\ \cmidrule(l){2-3} & 

        Loss of social connections  & Technology that makes it difficult to build, maintain, and nurture social connections. \\ \cmidrule(l){2-3} & 

        Apathy  & Feelings of apathy or powerlessness that arise from having no other options other than to accept the given conditions surrounding technology use. \\ \cmidrule(l){2-3} & 

        Surveillance  & Feelings of surveillance and invasion of privacy through the use of the technology. \\ \midrule

        Policies & 
        
        Funding & Policies, such as the [anonymized for review] fund, governing DOC funds---such as incoming funding sources, purpose of funds, or how funds are used to acquire technologies through contracts. \\ \cmidrule(l){2-3} & 

        Daily technology use  & Policies governing what, when, where, and how technology might be accessed or utilized by incarcerated individuals. 
        \\ \cmidrule(l){2-3} & 

        Communication  & Policies, such as ``\textit{Procunier vs. Martinez}'', that govern ways in which incarcerated individuals can communicate with loved ones outside. These policies dictate when mail or messages might be rejected, and the appeals process. \\ \cmidrule(l){2-3} & 

        Costs  & Policies that govern the cost of the technologies, including prices for phone calls, financial transactions, music, etc.\\ \midrule

        Stakeholders & 
        
        DOC staff & Any decision or action performed by DOC line officers, staff members, classification counselors, tier representatives, etc. \\ \cmidrule(l){2-3} & 

        Technicians  & Decision/action by technicians employed by the communication tech companies. \\ \cmidrule(l){2-3} & 

        Government officials & Any decisions or actions performed by state officials, lawmakers, or judicial officers. \\ \bottomrule
    \end{tabular}
    \caption{\textbf{Codebook}}
    \label{tab:codebook}
\end{table*}

\end{document}